%% file: main.tex
\documentclass[runningheads]{llncs}
\usepackage{amssymb}
\usepackage{amsmath}
\usepackage{graphicx}
\usepackage{url}
\usepackage[export]{adjustbox}
\usepackage{array}
\usepackage[labelformat=empty]{subfig}
\usepackage{color,soul} 
\soulregister\ref{7} 
\soulregister\cite{7} 
\usepackage{float} 
\usepackage{hyperref}
\hypersetup{colorlinks}
\usepackage[colorinlistoftodos]{todonotes}
\usepackage{booktabs} 
\usepackage{rotating}
\usepackage{multirow}
\usepackage{transparent}
\usepackage{tablefootnote}
\usepackage{algorithm}
\usepackage{algpseudocode}
\usepackage{subfig}
\usepackage{orcidlink}

\usepackage[subtle,margins=normal,leading=normal]{savetrees} 
\usepackage{bm} 
\usepackage{colortbl}
\usepackage{tabulary}

\begin{document}
\mainmatter              

\title{Federated Whole Prostate Segmentation in MRI with Personalized Neural Architectures}
\titlerunning{Federated Whole Prostate Segmentation}  %

\author{
Holger R. Roth\textsuperscript{\orcidlink{0000-0002-3662-8743}} \and
Dong Yang\textsuperscript{\orcidlink{0000-0002-5031-4337}} \and
Wenqi Li\textsuperscript{\orcidlink{0000-0003-1081-2830}} \and
Andriy Myronenko\textsuperscript{\orcidlink{0000-0001-8713-7031}} \and
Wentao Zhu\textsuperscript{\orcidlink{0000-0002-7505-9512}} \and
Ziyue Xu\textsuperscript{\orcidlink{0000-0002-5728-6869}} \and
Xiaosong Wang\textsuperscript{\orcidlink{0000-0002-3840-5658}} \and
Daguang Xu\textsuperscript{\orcidlink{0000-0002-4621-881X}}
}

\institute{
NVIDIA, Bethesda, MD, USA\thanks{Contact: \texttt{\{hroth,daguangx\}@nvidia.com}}
}
\authorrunning{***} 

\maketitle              

\begin{abstract}
Building robust deep learning-based models requires diverse training data, ideally from several sources. However, these datasets cannot be combined easily because of patient privacy concerns or regulatory hurdles, especially if medical data is involved. Federated learning (FL) is a way to train machine learning models without the need for centralized datasets. Each FL client trains on their local data while only sharing model parameters with a global server that aggregates the parameters from all clients. At the same time, each client's data can exhibit differences and inconsistencies due to the local variation in the patient population, imaging equipment, and acquisition protocols. Hence, the federated learned models should be able to adapt to the local particularities of a client's data. In this work, we combine FL with an AutoML technique based on local neural architecture search by training a ``supernet''. Furthermore, we propose an adaptation scheme to allow for personalized model architectures at each FL client's site. The proposed method is evaluated on four different datasets from 3D prostate MRI and shown to improve the local models' performance after adaptation through selecting an optimal path through the AutoML supernet.
\end{abstract}


\input{01_Introduction}

\input{02_Methods}

\input{03_Experiments}
\input{04_Conclusion}

\clearpage
\newpage
\bibliographystyle{splncs04}
\bibliography{bibliography}

\end{document}

%% file: 01_Introduction.tex

\section{Introduction}
\label{sec:intro}
The advancements of the last few years in medical image segmentation were dominated by deep learning (DL) approaches. DL mostly eliminated the need for handcrafting image features. However, it has been arguably replaced by the need of domain experts to design application-specific DL models. In particular, the medical image computing field has been dominated by popular hand-engineered network architectures such as 2D and 3D U-Net \cite{ronneberger2015u,cciccek20163d}, V-Net \cite{milletari2016v}, High-Res-Net \cite{li2017compactness}, DeepMedic \cite{kamnitsas2016deepmedic}, and many others. To get a good network design for a particular problem, one promising direction is to automate the time-consuming model designing process via AutoML techniques.
%
%
As another major challenge in model development, large amounts of data covering sufficient large range of examples are usually required to train accurate and robust models. To achieve this goal, hospitals and medical institutes often need to collaborate and host centralized databases for the development of clinical-grade DL models. This can become challenging due to data-privacy and various ethical concerns associated with data sharing in healthcare domain. One approach to combat such issues is through federated learning (FL), where only model and/or DL workflow parameters are shared among participating institutes instead of raw medical data.
Furthermore, it is well known that global robustness and local accuracy is in many cases conflicting: models trained on large centralized datasets might not always generalize well to the data at a particular imaging site due to various inconsistencies (scanner models, imaging protocols, patient populations, etc.) among the different sites. In this case, domain adaptation (DA) is often needed.
In this work, we propose to systematically tackle the three challenges in a unified framework: combining an FL algorithm with AutoML and the capability of global-local model adaptation. In particular, we implement a ``supernet'' training strategy that can be trained in a federated setting.
We believe AutoML and FL technologies are a natural fit for each other because of their complementary nature. By combining the two, we are also able to address the DA problem. 
For one, FL can circumvent the problem of hosting and accessing large centralized datasets by distributing the learning effort to several clients with their own local data. FL will only communicate the model gradients after a local round of training to a centralized server which aggregates the results and starts the next round of FL.
At the same time, AutoML with supernet design allows us to avoid hand-engineering of dataset-specific network architectures and a particular sub-network of the trained supernet can be used as a way of local domain adaptation to handle inconsistencies between the different contributing data sites. Next, we summarize related works.

\label{related_works}
\paragraph{\textbf{AutoML:}}
Recently, deep learning is applied for various applications, such as image recognition, semantic segmentation, object detection, natural image generation, etc.. However, for each specific task, particular network architectures often need to be hand-designed. Neural architecture search (NAS)~\cite{elsken2018neural} is one of the most common approaches to circumvent such hand-design of architectures in AutoML for DL applications. The goal of NAS is to automatically design neural network architectures without any human heuristics or assumptions. In addition to the model weights, after searching, the model architecture itself is optimized for the task at hand, while often still being generalizable to other datasets \cite{yu2019c2fnas}. 
A common concept in (one-shot) NAS and AutoML literature is the ``supernet'' 
\cite{liu2018darts,cai2018proxylessnas,you2020greedynas}. The main idea behind supernet is that we can create a large neural network including several candidate modules at each level of the networks. This supernet can be trained jointly, and from the supernet, specific sub-networks can be chosen by selecting a path through the module candidates. At deployment, the final architecture is selected from the supernet by assigning path weights to select particular module candidates.
Additional budgeting constraints, such as latency or number of model parameters, can be added to find optimal architectures for a given application. Recent works can achieve state-of-the-art results on computer vision tasks while being computationally efficient \cite{tan2019efficientnet}.
\paragraph{\textbf{Federated Learning:}}
FL enables collaborative and decentralized DL training without sharing raw patient data \cite{mcmahan2016communication}. Each client in FL trains locally on their own data and then submits their model parameters to a server that accumulates and aggregates the model updates from each client. Once a certain number of clients have submitted their updates, the aggregated model parameters are redistributed to the clients for local model update, and a new round of local training starts. While out of the scope of this work, FL can also be combined with additional privacy-preserving measures to avoid potential reconstruction of training data through model inversion if the model parameters would be leaked to an adversary \cite{li2019privacy}. Several works have shown the applicability of FL to medical imaging tasks \cite{sheller2018multi,li2019privacy,YANG2021101992}. Recent work that combines NAS approaches with FL has been proposed for the mobile phone applications \cite{zhu2020real}. As such, its focus is on reducing the computational requirements on the local edge devices, making its setting quite different from the ``cross-silo'' FL \cite{kairouz2019advances} medical image segmentation investigated here, where the focus is on model performance and personalization. The closest work in motivation to ours is \cite{he2020towards} which focuses on the non-I.I.D. setting but is restricted to using toy datasets for classification tasks, like CIFAR-10, and differs in its implementation details.
\paragraph{\textbf{Domain Adaptation:}}
Domain adaptation aims to tackle data inconsistencies among different domains, or between training data and unknown data. In its simplest form, fine-tuning, also known as transfer learning \cite{shin2016deep}, can help to adapt a pre-trained model to a particular target domain. 
More recent approaches for DA typically involve some form of adversarial learning to introduce a specific loss that can minimize the feature-level differences among different domains \cite{kamnitsas2017unsupervised} or through gradient back-propagation using adversarial training~\cite{ganin2015unsupervised,ganin2016domain}.
An alternative approach is coming from the ``image translation'' field where generative adversarial networks (GAN) are utilized to translate the image of one domain to mimic another domain. An important part of these approaches is the application of some form of cycle-consistency which is essential to train on un-paired data \cite{isola2017image,zhu2017unpaired,zhang2018task}
The common concept of adversarial training suggests that the gradients from external constraints will help balance various domains and change the model's feature representations.
\paragraph{\textbf{Contributions:}} Our proposed approach here is similar in that we will ultimately adapt the model's internal feature representations through the selection of an adapted sub-network of the trained supernet, but without the need to use computationally expensive adversarial learning schemes. Our contributions can be summarized as follows:
\begin{enumerate}
    \item We show that we can successfully train models through federated learning with comparable or better performance to models trained on centrally hosted data.
    \item We extent federated learning by introducing an AutoML approach for supernet model training.
    \item We show that finding an optimal path through the supernet can act as a form of local domain adaptation and bring performance gains for each individual client.
\end{enumerate}

%% file: 02_Methods.tex
\section{Method}
\label{sec:method}
\begin{figure}[htbp]
 \centering
    \subfloat[(a)]{
        \includegraphics[width=0.35\textwidth, valign=c]{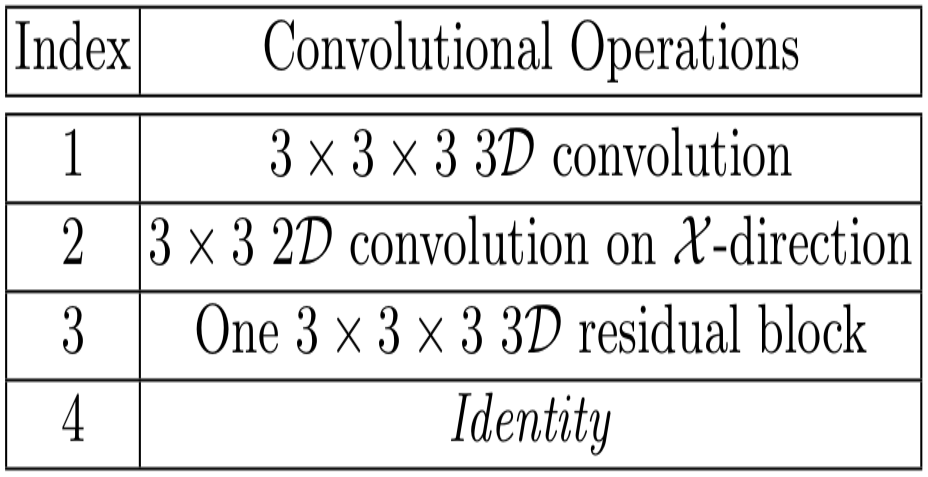}
        \label{fig:candidates}
    }%
    \\
    \subfloat[(b)]{
        \includegraphics[width=0.9\textwidth,valign=c]{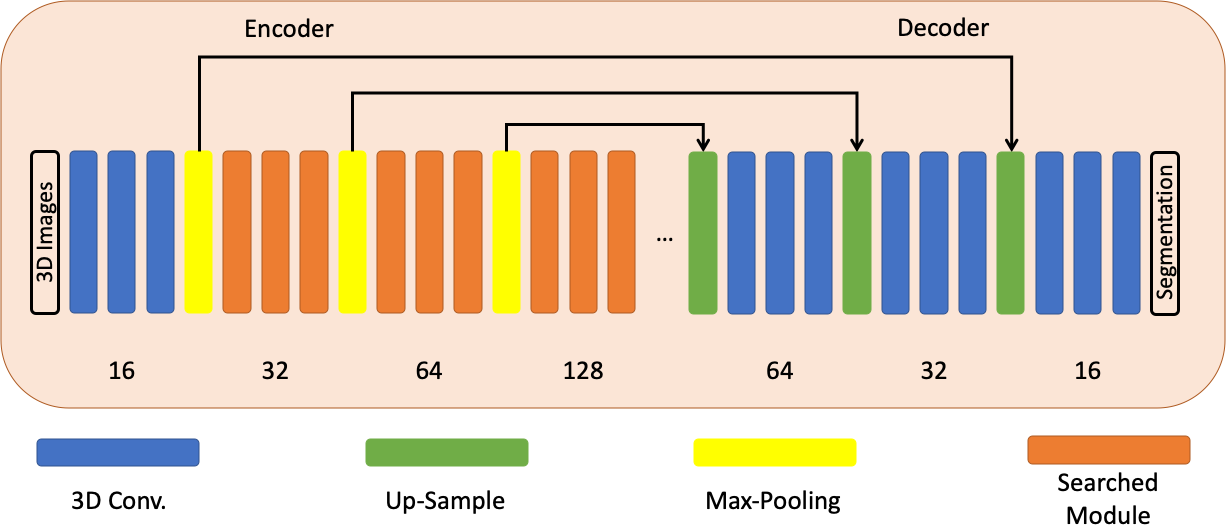} 
        \label{fig:supernet}
    }%
    \\
    \subfloat[(c)]{
        \includegraphics[width=0.6\textwidth,valign=c]{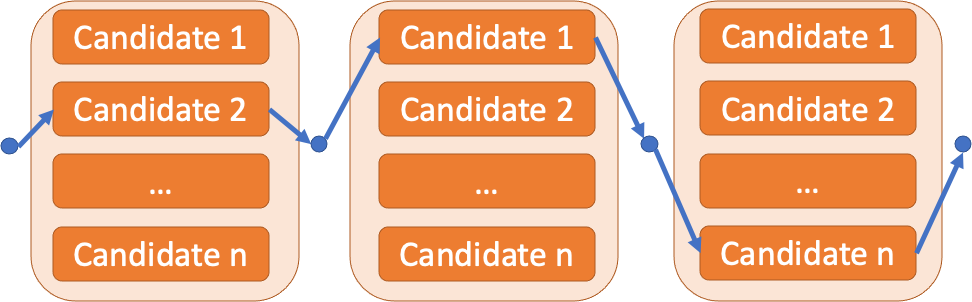}
        \label{fig:modules}
    }%
    \caption{(a) Searched network module candidates, (b) our supernet architecture $\mathcal{S}$, and (c) a potential path $m$ selecting several module candidates $\mathcal{M}$. Note, \textit{Identity} is only used if input/output sizes are the same.}%
\end{figure}

Here we describe the technical details of the FL and AutoML approach utilized in this work. 
The proposed method can be separated into two steps: 1) FL with AutoML supernet training and 2) local model adaptation by finding the best path through the supernet with respect to the local data. Both FL and AutoML procedures presented are designed for 3D medical image segmentation tasks.
\paragraph{\textbf{Client-Server-Based Federated Learning:}}
In its typical form, FL utilizes a client-server setup. Each client trains the same model architecture locally on their own data. Once a certain number of clients finished local training, the updated model weights (or their gradients) are sent to the server for aggregation. After aggregation, the new weights on the server are re-distributed to the clients to execute the next round of local model training. After several FL rounds, the models at each client are converged. Each client can be allowed to select their local best model by monitoring a certain performance metric on a local hold out validation set. In our experiments, we implement the \texttt{FederatedAveraging} algorithm proposed in \cite{mcmahan2016communication}. While there exist variants of this algorithm to address particular learning tasks, in its most general form, FL tries to minimize a global loss function $\mathcal{L}$ which can be a weighted combination of $K$ local losses $\{\mathcal{L}_k\}_{k=1}^{K}$ that each is computed on a client $k$'s local data.
Hence, FL can be formulated as the task of finding the model parameters $\phi$ that minimize $L$ given some local data $X$.
\begin{align}
\min_{\phi}\mathcal{L}(X; \phi) \quad \text{with} \quad  \mathcal{L}(X; \phi)=\sum_{k=1}^{K}w_{k}\;\mathcal{L}_{k}(X_{k}; \phi),
\label{eq:formalism}
\end{align}
where $w_k>0$ denote the weight coefficients for each client $k$, respectively. Note, that the local data $X_{i}$ is never shared among the different clients. Only the model weights are accumulated and aggregated on the server as shown in Algorithm~\ref{alg:fl}.
\begin{algorithm}[htb]
\caption{Client-server federated learning with \texttt{FederatedAveraging}~\cite{mcmahan2016communication,li2019privacy}. $T$ is the number of federated learning rounds and $n_k$ is the number of \texttt{LocalTraining} iterations minimizing the local loss $\mathcal{L}_{k}(X_{k}; \phi^{(t-1)})$ for a client $k$.}
\label{alg:fl}
\footnotesize
  \begin{algorithmic}[1]
    \Procedure{Federated Learning}{}
      \State{Initialize weights: $\phi^{(0)}$}
      \For{$t \gets 1\cdots T$}
        \For{$client\ k \gets 1\cdots K$}  \Comment{ \textit{Executed in parallel}}
        \State{Send $\phi^{(t-1)}$ to client $k$}
        \State{Receive $(\Delta \phi_k^{(t)}, n_k)$ from client's $\texttt{LocalTraining} (\phi^{(t-1)})$}
        \EndFor
        \State{$\phi_k^{(t)}\gets \phi^{(t-1)} + \Delta \phi_k^{(t)}$}
        \State{$\phi^{(t)}\gets \frac{1}{\sum_k{n_k}}\sum_k{(n_k\cdot \phi_k^{(t)})}$}
      \EndFor
      \State \Return $\phi^{(t)}$
    \EndProcedure
  \end{algorithmic}
\end{algorithm}
\paragraph{\textbf{AutoML with Supernet:}}
In order to allow for personalized neural architectures, we designed a supernet $\mathcal{S}$ consisting of various DL module candidates $\mathcal{M}$ suitable for 3D medical imaging tasks shown in Fig.~\ref{fig:candidates}. Each candidate $\mathcal{M}$ is a subgraph $s \in \mathcal{S}$, denoted as $\mathcal{N} \left ( s, w \right )$ with model weights $w$. These modules are optimized at multiple resolution levels to capture different levels of image features useful for the segmentation task. In general, we follow the popular encoder-decoder structure which has been successfully applied to many medical imaging tasks \cite{ronneberger2015u,cciccek20163d,milletari2016v} as shown in Fig. \ref{fig:supernet} with skip connections that concatenate features of the encoder with their corresponding layer in the decoder path.
During training, we choose arbitrary paths $m$ from the module candidates $\mathcal{M}$ following a uniform sampling scheme (see Fig. \ref{fig:modules}) to define a sub-network $s$ sampled from the supernet $\mathcal{S}$ as in Eq.~\ref{eq:train}.
\begin{equation}
\scriptsize
    W_{\mathcal{S}}=\underset{W}{\textup{argmax}}~\mathcal{L}\left ( \mathcal{N} \left ( \mathcal{S}, W \right ) \right )
    \label{eq:train}
\end{equation}
In this work, we choose the combination of Dice loss \cite{milletari2016v} and cross entropy loss as our loss function which is commonly used for segmentation tasks in medical imaging~\cite{isensee2021nnu}. Dice loss' major advantage is its ability to work well in segmentation tasks with an unbalance in the amount of foreground/background regions.
%
Once the supernet is trained, we can find a sub-network $s_0$ by identifying a locally optimal path through the supernet, effectively adapting the model to the target domain. During adaptation, the model parameters $\phi$ stay fixed and only the path weights are optimized for one epoch on the local validation set. This results in an optimal path $m_0 \in \mathcal{M}$ that defines our locally adapted sub-network $s_0 \in \mathcal{S}$ as Eq.~\ref{eq:val}.
\begin{equation}
\footnotesize
    s_0=\underset{s \in \mathcal{S}}{\textup{argmax}}~\mathcal{L}_{\textup{val}}\left ( \mathcal{N} \left ( s, w_s \right ) \right )
    \label{eq:val}
\end{equation}
%

%
%

%% file: 03_Experiments.tex
\section{Experiments \& Results}
\label{sec:experiments}
Our proposed method is evaluated on the task of 3D whole prostate segmentation in T2-weighted MRI. In particular, MRI has challenges of data inconsistencies due to variations in different imaging protocols and scanners used at each data contributing site, potentially causing drastic variations in contrast and intensity values. 

\paragraph{\textbf{Datasets:}}
We utilize prostate MRI datasets from four different publicly available data sources. 
MSD-Prostate\footnote{\url{http://medicaldecathlon.com}} \cite{simpson2019large}, PROMISE12 \footnote{\url{https://promise12.grand-challenge.org}} \cite{litjens2014evaluation}, NCI-ISBI13\footnote{\url{http://doi.org/10.7937/K9/TCIA.2015.zF0vlOPv}}, and ProstateX\footnote{\url{https://prostatex.grand-challenge.org}} \cite{litjens2014computer}.
For each dataset, we perform three random splits into training, validation, and testing sets at roughly 70\%, 10\%, and 20\% of the total number of cases of each dataset. The resulting number of cases for each dataset are shown in Table \ref{tab:results}. We average results across the testing splits of each random split. For reference, we show the results on a centralized dataset where all four datasets have been combined. We also compare the performance for models trained locally and through federated learning an each dataset's testing split. The performance of a standard 3D U-Net \cite{cciccek20163d} which is a subgraph of our supernet (when all candidates are type 1) is shown for a baseline comparison. We resample each image to a constant resolution of 0.5 mm $\times$ 0.5 mm $\times$ 1.0 mm and normalize all non-zero image intensities by subtracting their mean and dividing by their standard derivation on a per-image basis. 

\paragraph{\textbf{Implementation:}}
Both U-Net and the supernet are trained using randomly cropped patches of size $160\times 160\times 32$ from the input images and labels. We used a mini-batch size of 18 by selecting 3 random crops from any 6 random input image and label pairs. As the optimizer for training the supernet, we chose \textit{NovoGrad} which has typically faster convergence speed than the more commonly used Adam optimizer \cite{ginsburg2019stochastic}. The learning rate for supernet training was set to $1e^{-3}$. For finding the optimal path for the final sub-network we use the Adam optimizer with a learning rate of $1e^{-3}$.
Augmentation techniques like random intensity shifts, contrast adjustments, and adding Gaussian noise are applied during training to avoid overfitting to the training set. Our supernet has $3^3\times 4^6=110,592$ possible path combinations. Therefore, it is trained 10$\times$ longer than 3D U-Net to give it the opportunity to train most paths well. Both 3D U-Net baseline and the supernet are implemented with PyTorch\footnote{\url{https://pytorch.org}} using components from MONAI\footnote{\url{https://monai.io}} and NVFlare \footnote{\url{https://pypi.org/project/nvflare}} for FL communication. All models are trained on NVIDIA V100 GPUs with 16 GB memory. We monitor convergence on randomly chosen paths sampled from a uniform distribution during each validation to determine when the supernet is sufficiently trained across clients. The number of training iterations is chosen such that the likelihood of a path being selected during the entire training is at least >1. 

\begin{figure}[htbp]
\centering
    \includegraphics[width=1.0\textwidth]{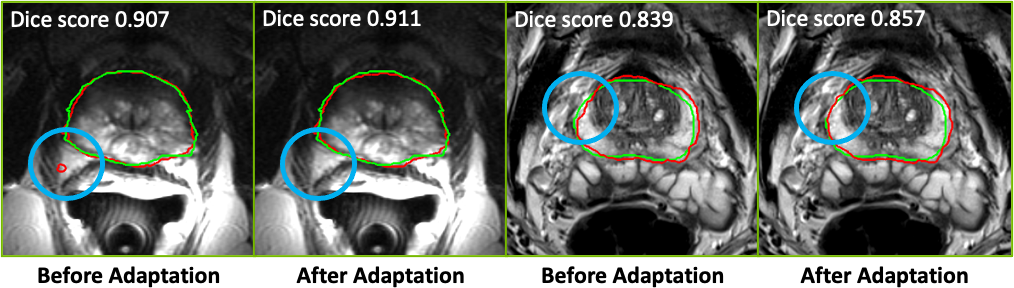}
    \caption{The visual comparison of results from supernet before and after adaptation. Left two figures show results from PROMISE12, and right two figures are from MSD. Green contours denote ground truth (boundaries), and red contours are predictions from neural networks. We can see that adaptation is not only able to remove false positives, but also capable to correct false negative regions (highlighted by circles).}
    \label{fig:vis}
\end{figure}

\paragraph{\textbf{Results:}}
Table~\ref{tab:results} show the performance for assuming local, centrally hosted, and federated datasets. Table~\ref{tab:gen} shows the better generalizibilty of supernet models trained in the FL. We show the performance of the proposed supernet training approach and its adaption to the local dataset distribution via path optimization, together with a baseline implementation of 3D U-Net using the same augmentation, optimization and hyperparameters to be comparable. Visualization of the results before and after model adaptation are shown in Fig.~\ref{fig:vis}. In descending order, most commonly chosen operations were 3D conv., 3D residual block, 2D conv., followed by identity.
\begin{table*}[htbp]
    \caption{Results for centralized dataset and each dataset trained locally and in federated learning. We show the performance of a baseline 3D U-Net and our proposed supernet (SN) approach. The average Dice of the local model's scores is shown (excluding the scores on centralized data). The highest scores for each dataset are marked in \textbf{bold}.}
    \label{tab:results}
      \centering
      \scriptsize
\begin{tabular}{lcccccc}
\hline\hline
\textbf{Cases}& \textbf{Central} & \textbf{NCI}& \textbf{PROMISE12}& \textbf{ProstateX}& \textbf{MSD} & \\
\hline
Training   & 172 & 45 & 35 & 69 & 23 \\
Validation & 23  & 6  & 5  & 9  & 3  \\
Testing    & 48  & 12 & 10 & 20 & 6  \\
\hline
Total      & 243 & 63 & 50 & 98 & 32 \\
\hline\hline
      &  &  &  &  &  \\
\hline\hline
\textbf{Avg. Dice [\%]}& \textbf{Central} & \textbf{NCI}& \textbf{PROMISE12}& \textbf{ProstateX}& \textbf{MSD} & \textbf{Avg. (loc.)}\\
\hline
U-Net (loc.)        & 89.72 &	90.15	&	83.59	&	90.73	&	86.17	&	87.66 \\
U-Net (fed.)        &  	    &	90.59	&	85.73	&	90.35	&	88.40	&	88.77 \\
\hline
SN (loc.)           & 90.11 &	90.42	&	83.51	&	90.76	&	86.76	&	87.86 \\
SN (fed.)           &       &	90.57	&	85.61	&	\textbf{91.07}	&	88.39	&	88.91 \\
\hline
SN (loc.) + adapt.  & \textbf{90.15} &	90.50	&	83.46	&	90.78	&	87.83	&	88.14 \\ 
SN (fed.) + adapt.  & 	    &	\textbf{90.68}   &	\textbf{86.15}   &	90.65   &	\textbf{88.74}   &	\textbf{89.06} \\
\hline\hline
\end{tabular}
\end{table*}
%
\begin{table}[htbp]
\caption{Generalizability of the supernet models trained locally SN (loc.), in federated learning \textbf{SN (fed.)}, and after local adaptation \textbf{SN (fed.) + adapt}. Here, client sites A, B, C, and D stand for NCI, PROMISE12, ProstateX, and MSD, respectively. We show the local average performance of the models (\textbf{Loc.}, same as in Tab.~\ref{tab:results}) and the generalizability (\textbf{Gen.}), i.e. the Dice score on other clients' test data. The total average generalizability score is denoted as $\mathbf{\overline{Gen.}}$ and the highest scores for each dataset are marked in \textbf{bold}.}
\label{tab:gen}
\scriptsize
\centering
\begin{tabular}{p{8mm}rrrr|r||rrrr|r||rrrr|r}
                             & \multicolumn{5}{c}{\textbf{SN (loc.)}}                                                                                                                                                                 & \multicolumn{5}{c}{\textbf{SN (fed.)}}                                                                                                                                                                      & \multicolumn{5}{c}{\textbf{SN (fed.) + adapt}}                                                                                                                                                         \\
                             & \multicolumn{4}{c}{\textit{\textbf{Test site}}}                                                                                                                    & \multicolumn{1}{l}{}              & \multicolumn{4}{c}{\textit{\textbf{Test site}}}                                                                                                                         & \multicolumn{1}{l}{}              & \multicolumn{4}{c}{\textit{\textbf{Test site}}}                                                                                                                    & \multicolumn{1}{l}{}              \\
\textit{\textbf{Train site}} & \multicolumn{1}{l}{\textbf{A}}             & \multicolumn{1}{l}{\textbf{B}}        & \multicolumn{1}{l}{\textbf{C}}        & \multicolumn{1}{l}{\textbf{D}}        & \multicolumn{1}{l}{\textbf{Gen.}} & \multicolumn{1}{l}{\textbf{A}}             & \multicolumn{1}{l}{\textbf{B}}        & \multicolumn{1}{l}{\textbf{C}}        & \multicolumn{1}{l}{\textbf{D}}             & \multicolumn{1}{l}{\textbf{Gen.}} & \multicolumn{1}{l}{\textbf{A}}             & \multicolumn{1}{l}{\textbf{B}}        & \multicolumn{1}{l}{\textbf{C}}        & \multicolumn{1}{l}{\textbf{D}}        & \multicolumn{1}{l}{\textbf{Gen.}} \\ \hline
\textbf{A}                            & \cellcolor[HTML]{EFEFEF}\textbf{90.4}      & \cellcolor[HTML]{DAE8FC}81.5          & \cellcolor[HTML]{DAE8FC}87.7          & \cellcolor[HTML]{DAE8FC}81.9          & 83.7                              & \cellcolor[HTML]{EFEFEF}\textbf{90.6}      & \cellcolor[HTML]{DAE8FC}86.1          & \cellcolor[HTML]{DAE8FC}90.9          & \cellcolor[HTML]{DAE8FC}85.5               & 87.5                              & \cellcolor[HTML]{EFEFEF}\textbf{90.7}      & \cellcolor[HTML]{DAE8FC}86.0          & \cellcolor[HTML]{DAE8FC}90.6          & \cellcolor[HTML]{DAE8FC}86.2          & 87.6                              \\
\textbf{B}                            & \cellcolor[HTML]{DAE8FC}84.6               & \cellcolor[HTML]{EFEFEF}\textbf{83.5} & \cellcolor[HTML]{DAE8FC}85.3          & \cellcolor[HTML]{DAE8FC}87.3          & 85.7                              & \cellcolor[HTML]{DAE8FC}87.4               & \cellcolor[HTML]{EFEFEF}85.6          & \cellcolor[HTML]{DAE8FC}89.7          & \cellcolor[HTML]{DAE8FC}88.3               & 88.5                              & \cellcolor[HTML]{DAE8FC}88.2               & \cellcolor[HTML]{EFEFEF}\textbf{86.2} & \cellcolor[HTML]{DAE8FC}90.3          & \cellcolor[HTML]{DAE8FC}\textbf{89.5} & 89.3                              \\
\textbf{C}                            & \cellcolor[HTML]{DAE8FC}84.9               & \cellcolor[HTML]{DAE8FC}72.8          & \cellcolor[HTML]{EFEFEF}\textbf{90.8} & \cellcolor[HTML]{DAE8FC}84.3          & 80.7                              & \cellcolor[HTML]{DAE8FC}89.7               & \cellcolor[HTML]{DAE8FC}\textbf{86.4} & \cellcolor[HTML]{EFEFEF}\textbf{91.1} & \cellcolor[HTML]{DAE8FC}87.7               & 87.9                              & \cellcolor[HTML]{DAE8FC}90.0               & \cellcolor[HTML]{DAE8FC}86.0          & \cellcolor[HTML]{EFEFEF}\textbf{90.7} & \cellcolor[HTML]{DAE8FC}87.7          & 87.9                              \\
\textbf{D}                            & \cellcolor[HTML]{DAE8FC}77.3               & \cellcolor[HTML]{DAE8FC}66.0          & \cellcolor[HTML]{DAE8FC}82.9          & \cellcolor[HTML]{EFEFEF}\textbf{86.8} & 75.4                              & \cellcolor[HTML]{DAE8FC}83.1               & \cellcolor[HTML]{DAE8FC}82.3          & \cellcolor[HTML]{DAE8FC}86.9         & \cellcolor[HTML]{EFEFEF}\textbf{88.4}      & 84.1                              & \cellcolor[HTML]{DAE8FC}84.5               & \cellcolor[HTML]{DAE8FC}83.9          & \cellcolor[HTML]{DAE8FC}87.3          & \cellcolor[HTML]{EFEFEF}88.7          & 85.2                              \\ \hline
\textbf{}                    & \multicolumn{1}{l}{\textit{\textbf{Loc.}}} & \textit{\textbf{87.9}}                & \multicolumn{1}{l}{}                  & \multicolumn{1}{l}{$\mathbf{\overline{Gen.}}$}     & \textit{81.4}                     & \multicolumn{1}{l}{\textit{\textbf{Loc.}}} & \textit{\textbf{88.9}}                & \multicolumn{1}{l}{}                  & \multicolumn{1}{l}{$\mathbf{\overline{Gen.}}$} & \textit{87.0}                     & \multicolumn{1}{l}{\textit{\textbf{Loc.}}} & \textit{\textbf{89.1}}                & \multicolumn{1}{l}{}                  & \multicolumn{1}{l}{$\mathbf{\overline{Gen.}}$}     & \textit{\textbf{87.5}}           
\end{tabular}
\end{table}

%% file: 04_Conclusion.tex
\section{Discussion \& Conclusions}
\label{sec:conclusions}
It can be observed from Table \ref{tab:results} that the supernet training with local adaptation in FL \textit{(SN (fed.) + adapt.)} achieves the highest average Dice score on the local datasets. At the same time,  the adapted models also show the best generalizability (see Table \ref{tab:gen}). This illustrates the viability of supernet training with local model adaption to the client's data. We furthermore observe a general improvement of the local supernet models' performance when trained in an FL setting versus local training. This means that in particular supernet model training can benefit from the larger effective training set size made available through FL without having to share any of the raw image data between clients. 
Overall, we achieve average Dice scores comparable to recent literature on whole prostate segmentation in MRI \cite{litjens2014computer,litjens2014evaluation,milletari2016v} and can likely be improved with more aggressive data augmentation schemes \cite{zhang2020generalizing,isensee2021nnu}.  
Further fine-tuning of the network weights (not the supernet path weights) is likely going to give performance boost on a local client but is also expected to reduce generalizability of the model. Methods of fine-tuning that do not reduce the robustness to other data sources (i.e. generalizability) gained through FL (e.g. learning without forgetting \cite{li2017learning}) is still an open research question and was deemed to be out of scope of this work.

In conclusion, we proposed to combine the advantages of both federated learning and AutoML. The two techniques are complementary and in combination, they allow for an implicit domain adaptation through the finding of locally optimal model architectures (sub-networks of the supernet) for a client's dataset. We showed that the performances of federated learning are comparable to the model's performance when the dataset is centrally hosted. After local adaptation via choosing the optimal path through the supernet, we can see an additional performance gain on the client's data. In the future, it could be explored if there is a set of optimal sub-networks that could act as an ensemble during inference to further improve performance and provide additional estimates such as model uncertainty. Furthermore, one could adaptively change the path frequencies used during supernet training based on sub-network architectures that work well on each client in order to reduce communication cost and speed-up training.
